# The Role of Permanent and Induced Electrostatic Dipole Moments for Schottky Barriers in Janus *MXY*/Graphene Heterostructures: a First Principles Study

*Yu-Qi Chen[1,2], Huan-Huan Zhang[1,2], Bo Wen[3], Xi-Bo Li[4], Yi-Feng Chai[1], Ying Xu[1], Xiao-Lin Wei[5], Wen-Jin Yin[1,2\*], and Gilberto Teobaldi[6,7]*

[1]*School of Physics and Electronic Science, Hunan University of Science and Technology, Xiangtan 411201, China*

[2]*Key Laboratory of Intelligent Sensors and Advanced Sensing Materials of Hunan Province, Hunan University of Science and Technology, Xiangtan 411201, China*

[3]*School of Physics and Electronics, Henan University, Kaifeng 475001, P. R. China*

[4]*Department of Physics, Jinan University Guangzhou 510632, P. R. China*

[5]*Department of Physics and Laboratory for Quantum Engineering and Micro-Nano Energy Technology, Xiangtan University, Xiangtan 411105, Hunan, China*

[6]*Scientific Computing Department, STFC UKRI, Rutherford Appleton Laboratory, Harwell Campus, OX11 0QX Didcot, United Kingdom*

[7]*School of Chemistry, University of Southampton, High Field, SO17 1BJ Southampton, United Kingdom*

## Abstract

The Schottky barrier height ($E_{SBH}$) is a crucial factor in determining the transport properties of semiconductor materials as it directly regulates the carrier mobility in opto-electronics devices. In principle, van der Waals (vdW) Janus heterostructures offer an appealing avenue to controlling the $E_{SBH}$. However, the underlying atomistic mechanisms are far from understood conclusively, which prompts for further research in the topic. To this end, here, we carry out an extensive first principles study of the electronic properties and $E_{SHB}$ of several vdW Janus *MXY/Graphene* (*M=Mo, W; X, Y=S, Se, Te*) heterostructures. The results of the simulations show that by changing the composition and geometry of the heterostructure's interface, it is possible to control its electrical contact, thence electron transport properties, from Ohmic to Schottky with nearly one order of magnitude variations in the $E_{SBH}$. Detailed analysis of the simulations enables rationalization of this highly attractive property on the basis of the interplay between the permanent dipole moment of the Janus *MXY* sheet and the induced one due to interfacial charge redistribution at the *MXY/Gr* interface. Such an interplay is shown to be highly effective in altering the electrostatic potential difference across the vdW Janus heterostructure, determining its $E_{SBH}$, thence Schottky (Ohmic) contact type. These computational findings contribute guidelines to control electrical contacts in Janus heterostructures towards rational design of electrical contacts in nanoscale devices.

## Introduction

The development of electronic devices towards smaller and smaller size and multifunction has become an attractive research field.[1-3] Owing to their unique features starting from controllable electronic properties, nano-materials have turned into a highly topical research subject. The successful experimental exfoliation of graphene opened for research in two-dimensional (2D) materials.[4] Over the years, it has been reported that graphene exhibits some excellent physical and chemical properties such as high carrier mobility[5] and high thermal conductivity[6], suggesting it as a potentially ideal nano-material to fabricate next-generation nano-electronic and optoelectronic devices. However, the lack of a sufficient large band gap has hindered its further practical application in these areas.[7] To overcome this limitation, different defect, doping, and structure tailoring approaches have been explored in order to open a band gap in graphene while preserving its pristine electron and heat transport properties. **Betti** *et al.* found that 1 nm wide armchair graphene nanoribbons have a band gap of 118 meV due to the occurrence of polaron formation.[8] However, it was also observed that the systems' carrier mobility drops significantly to less than 200 $cm^2V^{-1}s^{-1}$ for sub-10 nm graphene nanoribbons[9], with values as low as 60 $cm^2V^{-1}s^{-1}$ for in 1 nm wide armchair nanoribbons at room temperature.[10, 11]

In parallel to these studies, the search for graphene-analogue 2D materials with suitable band gaps and transport properties has expanded significantly.[12] $MoS_2$, a typical transition metal dichalcogenide (TMD) analogue of graphene, was successfully fabricated in 2010.[2, 13-16] Optical spectroscopy measurements indicate that the band gap of $MoS_2$ monolayer is about 1.8 eV[17], leading to an excellent on/off ratio in excess $10^8$.[18] Unfortunately, the carrier mobility of $MoS_2$ sheets is extremely low, in the 0.5-3 $cm^2V^{-1}s^{-1}$ range. Although structural modification can increase the carrier mobility to 200 $cm^2V^{-1}s^{-1}$,[19, 20] this value remains far from practical applications. Recently, by replacing the S-atoms on one side of $MoS_2$ by Se by a combination of plasma stripping and chemical vapor deposition, **Lu** *et al.* have succeeded in synthesizing Janus *MoSSe* sheets, which owns an intrinsic, electrostatic dipole moment.[21] Theoretical results have shown Janus *MoSSe* sheets have larger carrier mobility than $MoS_2$ monolayers. In detail, the carrier mobility of *MoSSe* monolayers can reach 157 $cm^2V^{-1}s^{-1}$ for holes, and 74 $cm^2V^{-1}s^{-1}$ for electrons. In addition, the bilayer or trilayer structures of *MoSSe* have an even higher electron and hole carrier mobility of 1194 $cm^2V^{-1}s^{-1}$ and 5894 $cm^2V^{-1}s^{-1}$, respectively as a result of the system's deformation potential.[22] These results have prompted for further investigations in alternatives Janus TMDs such as *WSeTe*, *MoSSe*, and *MoSTe*.[23-25] **Yang** *et al*. have predicted the electronic and optical properties of Janus *MoSeTe* in 2-H and 1-T phases by first-principles calculations. Their screened hybrid (HSE) results shows that the 2-H structure has a direct band gap of 1.86 eV, whereas the 1-T structure exhibits an indirect gap of about 0.39 eV.[26]

In addition to the approaches above, the study of van der Waals (vdW) heterostructures has also gained interest as promising avenues to nano-materials of function-tailored electronic properties. Existing studies in the field indicate that the assembly of vdW heterostuctures can not only change the band gap of the individual components, but also provide unique $E_{SBHS}$ and interface contact types. For example, it has been observed that by interfacing graphene with different semiconductors such as in *Graphene/MoS₂*[27-29], *Graphene/phosphorene*[30, 31], *Graphene/BN*[32], and *MoSSe/BN*[33] heterostructures, different electrical contacts can be engineered. Although these heterostructures do not lead to opening of a sufficiently large band gap for graphene, the interfacing does offer the opportunity to control the charge carrier transport properties of the composite system. Due to the quasi-metal properties of graphene, formation of Schottky barriers in graphene related heterostructures is facilitated. Recent studies have pointed out unique $E_{SBHS}$ and interface contact types in Janus *TMDs/Graphene* heterostructures, with the possibility of further tuning by strain or external electric fields.[34, 35] In spite of these advances, the atomistic mechanisms controlling the $E_{SBH}$ and charge-carrier properties in vdW Janus heterostructures are not fully understood, particularly with respect to the role of the internal electric field in the composite system.

To this end, the purpose of this work is to unveil and quantify the role of the internal electric field in determining the $E_{SBH}$ and interface contact properties of vdW Janus heterostructures. Here, we choose *MXY/Graphene (MXY/Gr, M=Mo, W; X, Y=S, Se, Te*) vdW Janus heterostructure as our case-studies, focusing on their geometry and electronic properties by first principles calculations. The results of the simulations show that the $E_{SBH}$ and interface contact can be controlled directly by altering the composition and interface geometry in the Janus *MXY/Gr* systems. This unique property originates from the synergistic interaction of the different dipole moments in the composites, where one dipole moment is provided by Janus *MXY* sheet and the other one is introduced by charge rearrangement of Janus heterostructure in response to the field from the MXY sheet. Based on the tunable direction and magnitude of these different dipole moments, the total induced electric field can be effectively regulated, resulting in controllable vacuum energy level difference ($\Delta V$) of the Janus heterostructure, which in turn affects the Schottky barrier height ($E_{SBH}$) and interface contact type.

**Computational Methods**

The first-principles calculations were performed by density functional theory (DFT) as implemented in the Vienna ab initio simulation program package (VASP).[36, 37] The Perdew-Burke-Ernzerh (PBE) exchange-correlation (XC) functional was employed to describe the exchange-correlation term within the generalized gradient approximation (GGA).[38, 39] The project augmented wave (PAW) was used to represent the atomic cores electronic density. The plane-wave energy cutoff was 500 eV, numerically checked to yield energies converged to within 0.001 eV/atom. The convergence criterion for the self-consistent DFT solution was set to $10^{-6}$ eV between two consecutive electronic steps. A 13×13×1 Monkhorst-Pack k-point grids was used for the single layer systems. Owing to the use of in-plane supercells of the individual components in the Janus heterostructure (vide infra), their Brillouin zone was sampled by means of a reduced 6×6×1 k-point grid. A vacuum buffer of 25 Å was applied to avoid interactions between adjacent images. All the atomic structures were fully relaxed until the residual atomic force were less than 0.001 eV/Å. vdW interactions were accounted for at DFT-D3 level.[40] It should be mentioned that Dipole corrections as implemented in the VASP programs have been used in all the calculations.

To investigate the relative stability of the Janus heterostructures, we used the binding energy per contact area ($E_b$), calculated as:

$$E_b = \frac{E_{MXY/Gr} - E_{MXY} - E_{Gr}}{A} \qquad (1)$$

where the $E_{MXY/Gr}$, $E_{MXY}$, and $E_{Gr}$ are the total DFT energy of the heterostructures and its individual (*MXY* and graphene sheets) components, respectively. *A* is the surface area of the heterostructure. For the sign convention used, the more negative $E_b$, the stronger and attractive the interactions between the components of the heterostructure.

## Results and Discussion

Both the single layer *MXY* (*M=Mo, W; X, Y=S, Se, Te*) JTMDs and graphene (*Gr*) are 2D structures with a honeycomb hexagonal lattice. The primitive cells for the two structures are shown in **Fig. 1(a)** and **1(b)**, respectively. We denote the corresponding lattice parameters $a_h$ and $b_h$ ($a_h=b_h$). Unlike graphene, there are three atomic layers in the *MXY*, with atoms *X* having a smaller atomic radius than atoms *Y*. **Table 1** summarizes the calculated basic structural and electronic properties for systems studied. It can been seen that the band gap of graphene is 0 eV with $a_h=b_h$ =2.47 Å. All the Janus *MXY* structures are semiconductors with band gaps ranging from 1.17 eV (*MoSTe*) to 1.70 eV (*WSSe*). The $a_h=b_h$ lattice parameters vary from 3.15 Å (MoS$_2$) to 3.43 Å (*WSeTe*). The calculated intrinsic electrostatic dipole moment for the individual *MXY* system range from 0.036 e·Å (*WSSe*) to 0.076 e·Å (*WSTe*). Such an intrinsic electrostatic dipole moment results in a difference in work function *(ΔV)* between the two sides of the *MXY* sheet, with larger *ΔV* values being associated with larger electrostatic dipole moments (*μ*).

**Table 1.** *The calculated lattice parameters ($a_h$, $b_h$ Fig. 1), M-X bond length ($D_{M-X}$), bond angle ($\angle XMY$), band gap ($E_g$), work function difference ($\Delta V$), and intrinsic dipole moment ($\mu$) for the considered MXY systems, together with reference MoS$_2$ and graphene data. All the results have been calculated in primitive cells.*

| System | $a_h=b_h$ (Å) | $D_{M-X}$ (Å) | $D_{M-Y}$ (Å) | $\angle XMY$ ($^0$) | $E_g$ (eV) | $\Delta V$ (eV) | $\mu$ (e·Å) |
|---|---|---|---|---|---|---|---|
| *MoS$_2$* | 3.15 | 2.41 | 2.41 | 81.97 | 1.75 | 0 | 0 |
| *MoSSe* | 3.21 | 2.41 | 2.53 | 82.27 | 1.67 | 0.78 | 0.037 |
| *MoSTe* | 3.32 | 2.42 | 2.71 | 82.49 | 1.17 | 1.48 | 0.078 |
| *MoSeTe* | 3.39 | 2.54 | 2.71 | 83.35 | 1.36 | 0.74 | 0.041 |
| *WSSe* | 3.25 | 2.42 | 2.54 | 81.72 | 1.70 | 0.71 | 0.036 |
| *WSTe* | 3.36 | 2.44 | 2.72 | 81.84 | 1.23 | 1.4 | 0.076 |
| *WSeTe* | 3.43 | 2.56 | 2.73 | 82.64 | 1.35 | 0.69 | 0.039 |
| *Graphene* | 2.48 | 1.42 | 1.42 | 120 | 0 | 0 | 0 |

The vdW Janus heterostructures have been initially constructed based on the individually relaxed *MXY* and graphene components, and eventually relaxed in the composite configuration. Because of the lattice mismatch between the individual components (**Table 1**), a 4×4×1 supercell of graphene and a 3×3×1 supercell of *MXY* are used to build the heterostructures. As for *MoS$_2$*, *MoSSe*, and *MoSTe*, graphene part undergoes a little strain force, while the graphene will surfer from a little tensile force for the rest structures. However, all the heterostructures lead to a contained lattice mismatches systematically lower than 5% can be achieved. Due to the asymmetry of the Janus MXY sheets, two types of interface modes can be formed, which we denote *Mode-I* and *Mode-II*. In *Mode-I*, the *X*-atoms with smaller radius points to graphene sheet (**Fig. 1(c)**). We label this geometry as *YMX/Gr* in the following. Conversely, in *Mode-II* the larger Y-atoms face the graphene sheet (**Fig. 1(d)**), leading to a *XMY/Gr* notation.

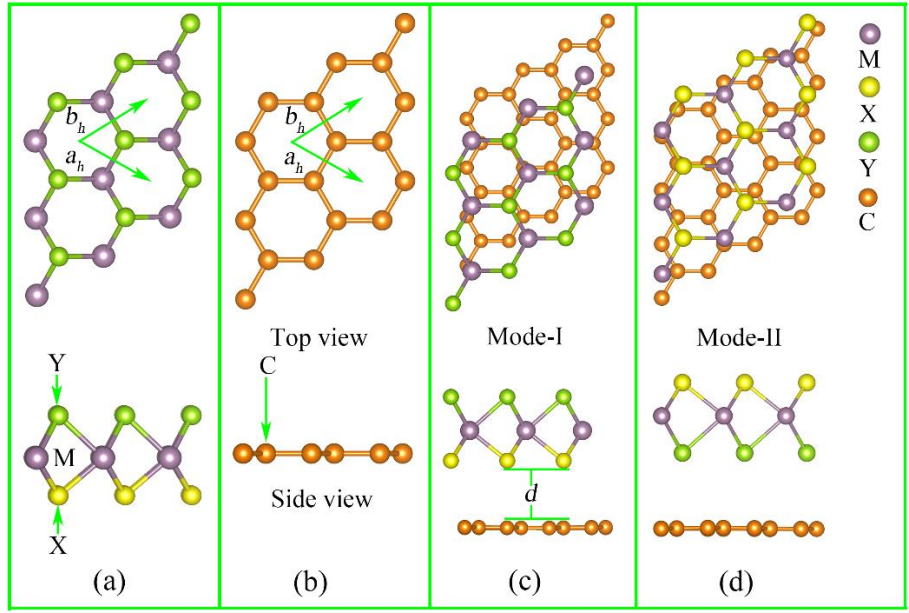

***Figure 1. (Color online)*** *The atomic structure for (a) the Janus TMDs MXY monolayer, (b) the graphene monolayer, and the different MXY/graphene vdW heterostructures geometries considered: (c) Mode-I, (d) Mode-II. The top and side view of the structure is shown in the upper and lower part of each panel. The C, M, X, and Y atoms are shown as orange, mauve, yellow, and green spheres, respectively. The symbol "d" marks the shortest distance between the MXY and Graphene sheet.*

**Table 2** reports the calculated binding energy ($E_b$), interlayer distance ($d$), mismatch ratio ($\delta$), Schottky barrier height ($E_{SBH}$), and dipole moment ($\mu$) of the *MXY/Gr* (*M=Mo, W; X, Y=S, Se, Te*) systems in different interface modes (I and II). $\delta$ for these heterostructures is systematically smaller than 5%, suggesting contained biases due to it. To check their stability, we examine $E_b$ for these heterostructures. Interestingly, all the $E_b$ values for the different *MXY/Gr* systems are significantly more negative by a factor of roughly 2 or more than for *MoS₂/Gr* (-6.86 meV/Å²), pointing to a larger stability of the *MXY/Gr* composites with respect to *MoS₂/Gr*. Furthermore, we can find that the $E_b$ of *Mode-II* is lower than that of *Mode-II*, indicating an energy favorability of the former over the latter. Surprisingly, it can be seen in **Table 2** that the most stable heterostructure in *Mode-II* results in larger interlayer distance ($d$). This trend is in stark contrast with the results for other Janus heterostructures such as *MoSSe/GaN*.[28] However, the differences in interlayer distances between *Mode-I* and *Mode-II* are qualitatively consistent with previous results suggesting that interfacing via the *X* atom side leads to smaller *d* values than in interfaces with the *Y* side, consistent with the reduced atomic radius going from X to Y atoms.

**Table 2.** *The calculated bind energy (E_b), interlayer distance (d), mismatch ratio (δ), Schottky barrier height (E_{SHB}), difference between the vacuum level on either side of the system (ΔV_v), and dipole moment (μ) of the Janus MXY/Gr heterostructures (M=Mo, W; X, Y=S, Se, Te) in different interface contact modes (I and II, Fig. 1). Depending on the different interface contact type, the Schottky barrier can be classified into p-type and n-type.*

| Mode | Mode | $D$ (Å) | $\delta$ (%) | $E_b$ (meV/Å²) | $E_{SHB}$ (eV) | | $\Delta V$ (eV) | $\mu$ (e·Å) |
|---|---|---|---|---|---|---|---|---|
| | | | | | $\Phi_p$ | $\Phi_h$ | | |
| *MoSSe/Gr* | S-C (I) | 3.36 | 2.50 | -16.29 | 1.35 | **0.02** | -0.385 | 0.178 |
| | Se-C (II) | 3.45 | | -17.00 | 0.80 | **0.60** | **+1.00** | 0.461 |
| *MoSTe/Gr* | S-C (I) | **3.35** | 2.00 | -18.78 | 1.17 | **0.00** | -1.11 | 0.519 |
| | Te-C (II) | 3.60 | | **-21.33** | **0.10** | 1.15 | **+1.65** | **0.772** |
| *MoSeTe/Gr* | Se-C (I) | 3.45 | 2.90 | -14.98 | 0.75 | **0.50** | -0.49 | 0.234 |
| | Te-C (II) | 3.60 | | -16.17 | **0.15** | 1.10 | **+0.94** | 0.447 |
| *WSSe/Gr* | S-C (I) | 3.50 | **1.30** | -16.88 | 1.25 | **0.25** | -0.48 | 0.222 |
| | Se-C (II) | 3.43 | | -17.56 | **0.70** | 0.90 | **+0.95** | 0.436 |
| *WSTe/Gr* | S-C (I) | 3.35 | 4.00 | -18.39 | 1.15 | **0.20** | -1.20 | 0.248 |
| | Te-C (II) | 3.60 | | **-20.37** | **0.00** | 1.25 | **+1.50** | 0.308 |
| *WSeTe/Gr* | Se-C (I) | 3.45 | 1.90 | -11.54 | **0.55** | 0.75 | -0.49 | 0.102 |
| | Te-C (II) | 3.60 | | -12.81 | **0.20** | 1.20 | **+0.82** | 0.170 |
| *MoS₂/Gr* | S-C | 3.35 | 4.50 | -6.86 | 1.10 | **0.02** | 0.46 | 0.210 |

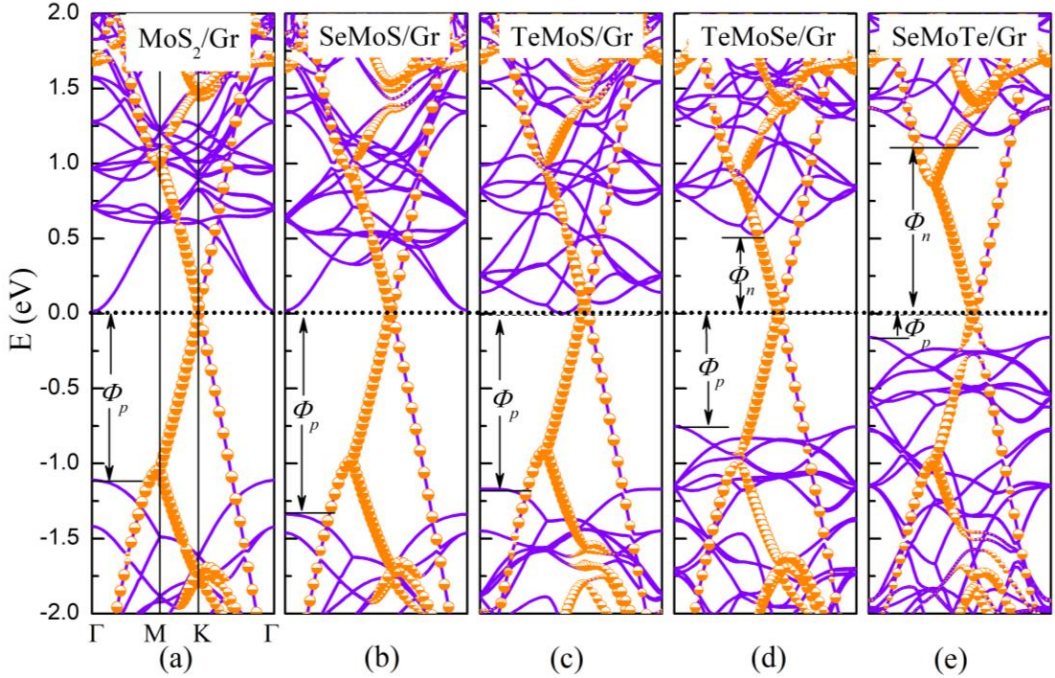

*Figure 2.* (*Color online*) *The atomic projected weighted band structure of the vdW Janus heterostructures (a) MoS₂/Gr, (b) SeMoS/Gr in Mode-I, (c) SMoSe/Gr in Mode-II, (d) TeMoS/Gr in Mode-I, and (e) SMoTe/Gr in Mode-II. The purple and orange bands are primarily due to the Janus MXY monolayer and graphene sheet, respectively. The Fermi level is set at 0 eV.*

**Fig. 2** reports the calculated band structure for the vdW Janus *MXY*/*Gr* heterostructures and *MoS₂/Gr* for comparison. It can be seen that some vdW Janus heterostructures are semiconductors with small band gaps. For example, the band gap of *MoSSe/Gr* is 57 meV in *Mode-I* and 61 meV in *Mode-II*, suggesting that interfacing *MoSSe* and Gr is effective in opening a finite band gap in the graphene sheet. As for the MoSSe side of *MoSSe/Gr*, its band gap measured from the Valence Band Maximum (VBM) to the Conduction Band Minimum (CBM) is reduced with respect to isolated *MoSSe* (1.67 eV, **Table 1**), being about 1.35 eV in *Mode-I* and 1.4 eV in *Mode-II*. In addition, it can be found that the energy difference from the VBM to the Fermi level ($E_f$) is different from the CBM-$E_f$ gap. As elaborated in the following, the asymmetry of these energy difference reflects the type (e.g. Schottky vs. Ohmic) and barrier of the interfacial electrical contact.

According to the Schottky-Mott rule for semiconductor-metal interface contacts,[35, 41] p-type Schottky barrier can be described by the energy difference between the VBM of *MXY* and $E_f$ of graphene. Conversely, the n-type Schottky barrier is defined as the energy difference between the CBM of *MXY* and $E_F$. Formally, the Schottky barriers for n-type ($\Phi_n$) and p-type ($\Phi_n$) of the heterostructure can written as:[35, 41]

$$\Phi_n = W_{Gr} + \Delta V_{MXY/Gr} - \chi_{MXY} \qquad (2)$$

$$\Phi_p = I_{MXY} - W_{Gr} - \Delta V_{MXY/Gr} \qquad (3)$$

$W_{Gr}$ is the work function of the graphene sheet. $\Delta V$ is the difference in the energy of the vacuum level above the *MXY* and *Gr* sides of the heterostructure. $\chi_{MXY}$, and $I_{MXY}$ indicate the electron affinity and ionization potential of the *MXY/Gr* heterostructure, respectively.

**Fig. 3** reports the calculated Schottky barrier heights for the vdW Janus *MXY/Gr* and *MoS₂/Gr* heterostructures. Simulations of *MoS₂/Gr* results in $\Phi_n$=0.02 eV and $\Phi_p$=1.1 eV, suggesting this heterostructure has an n-type Schottky contact owing to the lower $\Phi_n$ value by comparison to $\Phi_p$. The results for the Janus *MXY/Gr* heterostructures are substantially different and strongly dependent on the interface mode. For *MoSSe/Gr* in *Mode-I* (**Fig. 3(a)**) $\Phi_p$=1.35 eV and $\Phi_n$=0.02 eV, suggesting a small-barrier n-type Schottky contact. However, the results of the simulations clearly indicate that as the composition of the *MXY* subsystem (*M=Mo, W; X, Y=S, Se, Te*) changes, both the contact type and $E_{SBH}$ can be changed. Specifically, *MoSTe/Gr* (*Mode-I*) has an n-type Ohmic contact with $\Phi_n$=0 eV. In contrast, *MoSeTe/Gr*, *WSSe/Gr* and *WSTe/Gr* form n-type Schottky contacts with 0.2 eV <$\Phi_n$<0.5 eV. Interestingly, *WSeTe/Gr* (*Mode-I*) result in a p-type Schottky contact with $\Phi_p$=0.55 eV that can be reduced to as much as 0 eV for a *Mode-II* interface geometry, leading to an Ohmic contact. Similar results are found across the different composition studied. For example, the $E_{SBH}$ of *MoSSe/Gr* increases from $\Phi_n$=0.0 eV in *Mode-I* to $\Phi_n$=0.6 eV in *Mode-II*. As for *MoSTe/Gr*, *MoSeTe*, *WSSe* and *WSTe*, transition from *Mode-I* to *Mode-II* geometries changes the Schottky contact from n-type to p-type. These results indicate that, by adjusting the composition and interface geometry of vdW Janus heterostructures, it is possible to control the type (Schottky or Ohmic) of the electrical contact and, for Schottky one, tune the corresponding barrier by nearly one order of magnitude e.g. 0.1 <$\Phi_p$< 0.7 eV and 0.2 <$\Phi_n$< 0.6 eV. The known limitations of the used PBE XC-functional in underestimating band-gaps prompts for validation of the accuracy of the present results. To this end, we reconsidered the vdW Janus *SeMoS/Gr* heterostructure in *Mode-I* at HSE06 level.[42] As seen in the Supporting Information (**SI-Fig.1**) the band gap of the MoSSe part of vdW Janus heterostructure is increased by 0.4 eV with respect to the PBE results. However, the Ohmic type of the interface ($\Phi_n = 0$ eV) remains

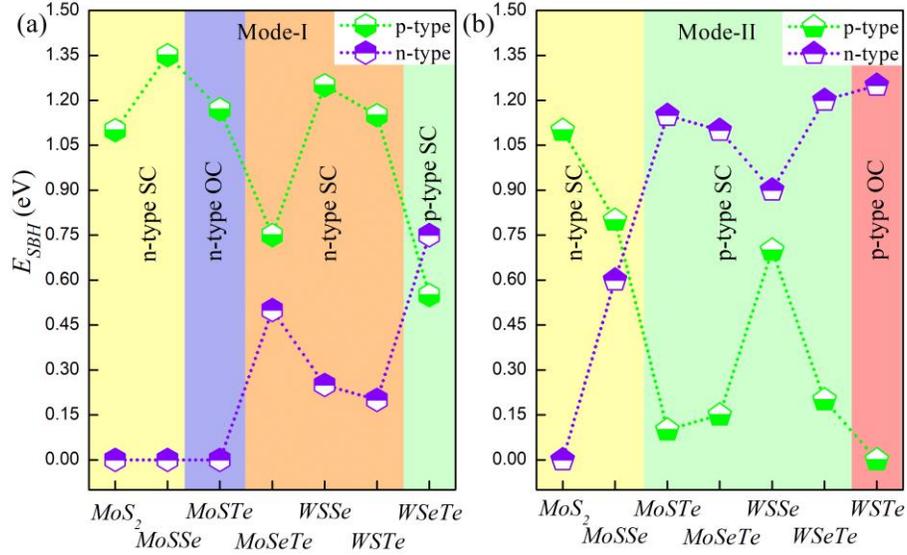

***Figure 3. (Color online)*** *The $E_{SBH}$ of the MXY/Gr (M=Mo, W; X, Y=S, Se, Te), and MoS$_2$/Gr heterostructures as a function of the interface contact geometry: (a) Mode-I, and (b) Mode-II. The results for p-type and n-type Schottky contacts are shown in green and purple, respectively.*

The above result demonstrates that the composition and interface geometry in vdW Janus heterostructure *MXY/Gr* can not only change the type of electrical contact between the constituents, but also control the magnitude of the Schottky barrier height. In the following we clarify the atomistic mechanisms of such effects. Although no dipole moment exists on the individual *MoS$_2$* or graphene sheet, as seen in **Table 2**, the vacuum energy level on either side of the *MoS$_2$/Gr* heterostructure is different (lower on the graphene side and higher on the MoS$_2$ side). As shown in **Fig. 4(a)**, this results stems from interfacial electron redistribution, with depletion (accumulation) of electronic charge on the graphene *(MoS$_2$)* side of the heterostructure. As a result of this interfacial electron rearrangement and ensuing formation of an interfacial permanent dipole (0.21 *e·Å*) the heterostructure present an in-built electrostatic field, leading to $\Phi_p$ =1.1 eV and $\Phi_n$ =0 eV. In contrast to the MoS$_2$ sheet, an intrinsic dipole moment (perpendicular to plane) exists for the individual *MXY* monolayers (**Table 1**). Thus, as *MXY* and graphene are interfaced in the vdW Janus heterostructure, both the dipole moments of *MXY* and *MXY/Gr* may contribute to the total induced electric field in the overall system. The vacuum energy level difference ($\Delta V$) between the two sides of the vdW Janus MXY/Gr heterostructure originates from the synergistic effects of the intrinsic dipole (*MXY*) and charge-reorganization induced electric field (*MXY/Gr*). The calculated results in **Table 2** show that *Mode-II* leads to larger $\Delta V$ and dipole moment ($\mu$) than *Mode-I*, suggesting the presence of larger induced electric field in *Mode-II*. Further analysis reveal that the direction of the electric field in *Mode-I* and *Mode-II* interfaces is totally different, owing to different interfacial charge redistribution. Taking *MoSeTe/Gr* as an example, it can be found that the vacuum energy level on the *MoSTe* side in *Mode-I* is higher than that on the graphene side (**Fig. 4(b)**). This is opposite than what calculated for the same system in *Mode-II* (**Fig. 4(c)**). <mark>In order to rule out possible depolarization artefacts due to periodic boundary condition (PBC) treatment of one individual vdW Janus structure, the simulations for *TeWS/Gr* were repeated for a different set-up consisting of two back-to-back heterostructures separated by a vacuum buffer of 20 Å. As seen in the Supporting Information (**SI-Fig. 2**), the depolarization can be as large as 30% for an individual *MoSSe* sheet, while the difference in calculated *ΔV* for the two *TeWS/Gr* set-ups is 0.01 eV, which demonstrates an effectively negligible role for PBC-induced depolarization artefacts for the single-heterostructure simulation cells used throughout the manuscript.</mark>

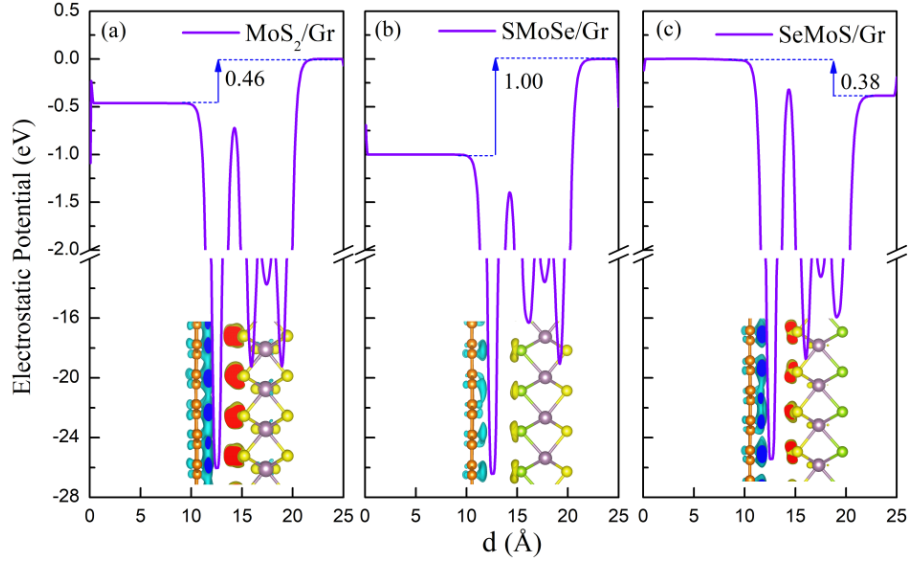

***Figure 4.*** *Comparison between the planar average of the calculated electrostatic potential and vacuum-plateaus differences for (a) MoS/Gr, (b) SMoSe/Ge and (c) SeMoS/G heterostructures. The highest vacuum-plateau has been set to zero. The C, Mo, S, and Se atoms are shown as orange, mauve, yellow, and green spheres, respectively.* *The color part (inset) represents the electronic density difference calculated by total charge density minus individual isolated part of heterostrure. The blue and cyan part means the depletion of electronic charge on graphene side, while red and yellow part denotes the accumulation of electronic charge on MXY side of heterostructure.*

In order to explain the changes in interface contact, we propose the polarization model in **Fig. 5**. The total electric field across the heterostructure (**$E_{total}$**) can be written as the sum of the electric fields due to the intrinsic *MXY* electrostatic dipole (**$E_{MXY}$**) and the electrostatic dipole induced by the MXY-Gr charge-redistribution (**$E_{induced}$**):

$$\vec{E}_{total} = \vec{E}_{MXY} + \vec{E}_{induced} \tag{4}$$

In *Mode-I*, the direction of the *MXY* dipole moment is opposite to that of the one induced by the *MXY/Gr* charge-redistribution, resulting in different vacuum energy level on graphene (higher) and *MXY* (lower) sides. The opposite direction of the *MXY* and induced dipole moments and associated electrostatic fields (**$E_{MXY}$** and **$E_{induced}$**) results in a reduced **$E_{total}$** in *Mode-I*. For this case, **$E_{total}$** goes from the *MXY* to the graphene sheet. Conversely, in *Mode-II*, the *MXY* and induced *MXR/Gr* dipole moments have the same direction (from *Gr* to *MXY*), which results in an enhancement of **$E_{total}$** and *ΔV*. The composition- and geometry-tunable magnitude and direction of $E_{total}$ (*ΔV*) as sum of the *MXY* and *MXY/Gr* contributions is responsible for the variety of the types of electrical contact and associated barriers seen in **Table 2** and **Figures 3-4**.

Considering *Mode-I* in **Fig. 5(a)**, *MoSSe/Gr* yields a small barrier ($Φ_n$=0.02 eV) n-type Schottky contact with $Φ_n$=0.02 eV. Turning to *MoSTe/Gr*, its $E_{total}$ and *ΔV* are larger than that of *MoSSe/Gr* (**Table 2**) due to the larger $E_{MoSTe}$ by comparison to $E_{MoSSe}$ (**Table 1**). The 0.025 eV difference between the ΔV (**Table 1**) and *ΔV* (**Table 2**) values for the isolated *MoSTe* and *MoSSe*, and the composite *MoSTe/Gr* and *MoSSe/Gr* systems indicate a negligible role for the interface-specific (*Mode-I* vs. *Mode-II*) $E_{induced}$. In turn, the enhanced $E_{total}$ pushes the CBM of *MoSTe* to cross the Fermi level resulting in an n-type Ohmic contact for *MoSTe/Gr*.

As seen in **Fig. 5(a)** and **5(b)**, due to the synergistic effects of the *MXY* and *MXY/Gr* electrostatic dipole moments, *Mode-I* and *Mode-II* interface models results in opposite directions for $E_{total}$. Taking *MoSeTe/Gr* as an example, $E_{total}$ in *Mode-I* points from *MoSeTe* to *graphene*. Conversely, in *Mode-II*, and due to the inversion of $E_{MXY}$, $E_{total}$ points from *graphene* to *MoSTe*. This will shift $E_f$ towards the VBM of *MoSTe*, resulting in a p-type Schottky contact. Therefore, the additive potential step $ΔV_v$ (**Table 2**) associated with the $E_{total}$ is shown to be highly effective in regulating the Fermi level (of graphene) and the VBM of *MXY*, resulting in a tunable electrical contact at the interface.

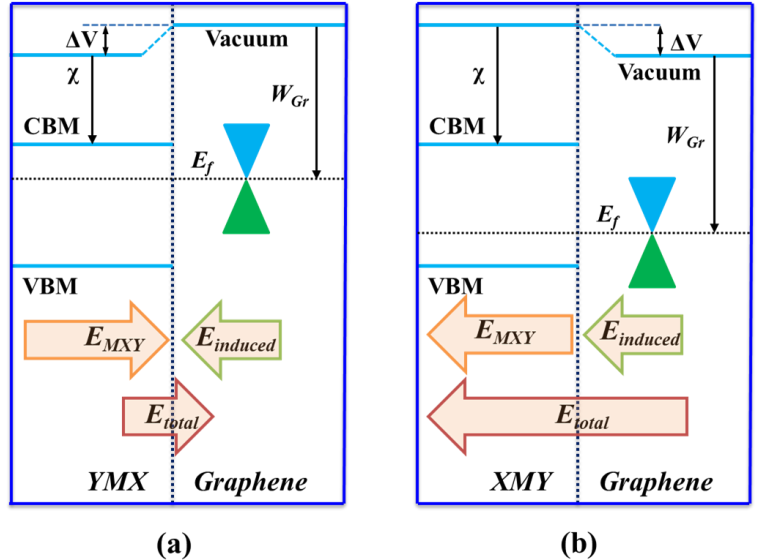

**(a)**                    **(b)**

*Figure 5. (Color online) The polarization model for the Janus MXY/Graphene heterostructure in (a) Mode-I and (b) Mode-II interface contact geometry. $E_f$: Femi level. The total, intrinsic MXY, and induced MXY/Gr. electrostatic fields are indicated as $E_{total}$, $E_{MXY}$, and $E_{induced}$, respectively. The conduction band minimum and valence band maximum are indicated as CBM and VBM, respectively. $W_{Gr}$: graphene work function. ΔV: the difference between the vacuum levels on the MXY and graphene side of the heterostructure.*

The previous qualitative analysis of the effects of *ΔV* and $E_{total}$ on the $E_{SBH}$ can be made more quantitative on the basis of Poisson equation:[43-45]

$$\nabla^2 V(z) = -\frac{\rho(z)}{\varepsilon_0} \tag{5}$$

$$E_{total}(z) = -\nabla V(z) \tag{6}$$

where $\rho(z)$ and $V(z)$ correspond to the (plane-averaged) total charge density and associated electrostatic potential along the non-periodic direction ($z$) of the *MXY/Gr* heterostructure, and $\varepsilon_0$ is the vacuum permittivity, respectively.

**Fig. 6** reports the calculated $E_{total}$ and $E_{MXY}$ based on Eqs. 5-6 for different representative cases of the considered vdW Janus heterostructures. In **Fig. 6(a)**, it can be found that *MoSSe/Gr* in *Mode-II* generates the $E_{total}$ of 0.27 eV/Å, $E_{MXY}$ of 0.178 eV/Å, and $E_{induced}$ of 0.09 eV/Å, with all the electric fields pointing from the graphene to the *MXY* monolayer. Owing to the larger dipole of *SMoTe* by comparison to *SMoSe* (**Table 1**), changing *SMoSe/Gr* (**Fig. 6a**) for *SMoTe/Gr* (**Fig. 6b**) results in larger $E_{total}$ at about 0.314 eV/Å and $\Delta V$ (**Table 2**). As a result, the $E_f$ will be pushed downwards and cross the middle of the band gap, generating a p-type Schottky contact of *SMoTe/Gr* in contrast to the n-type Schottky contact for *SMoSe/Gr*.

$E_{total}$ can also be used to explain the change of electrical contact as a function of the *Mode-I* or *Mode-II* interface geometry. For this analysis, we choose *MoSeTe/Gr* as a case-study. As shown in **Fig. 6(c)**, for *MoSeTe/Gr* in *Mode-II*, an $E_{total}$ of 0.18 eV/Å is generated with the direction pointing from graphene to *MoSeTe* and formation of a p-type Schottky contact. Conversely, in a *Mode-I* geometry, this leads to an overall negative $E_{total}$ at about -0.09 eV/Å on the graphene sheet (pointing from the *MoSeTe* side to the vacuum-exposed side of graphene) and inversion of the sign of $\Delta V$ with respect to the value for the *Mode-II* geometry (**Table II**), which in turn originates an n-type Schottky contact.

Overall, these results reiterate that the synergistic effects due to both the intrinsic electrostatic dipole of the Janus *MXY* component and the geometry-dependent, induced one due to charge redistribution at the *MXY/Gr* interface can be used to very effectively tune the type of interface contact and its $E_{SBH}$ in Janus vdW *MXY/Gr* heterostructures.

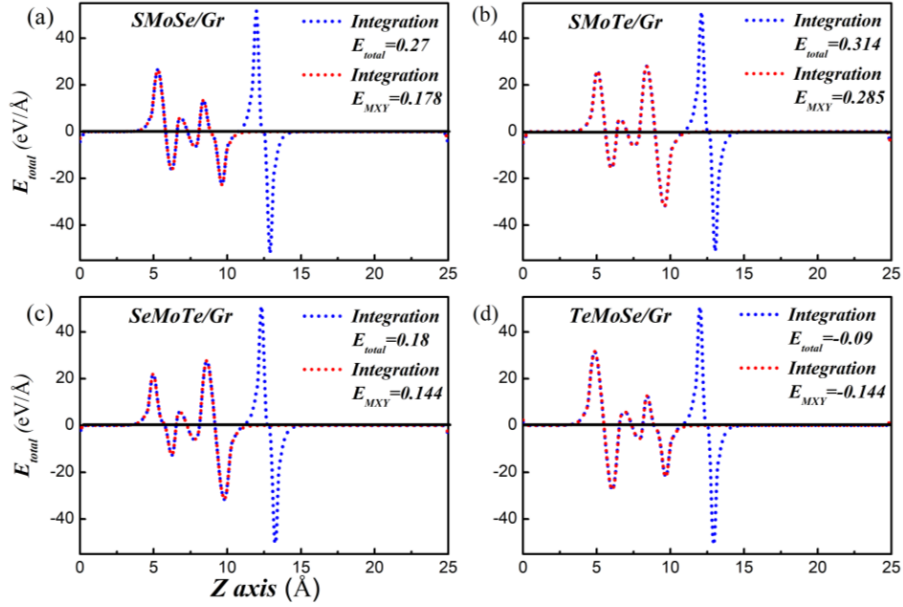

***Figure 6. (Color online)*** *The total electric field ($E_{total}$, blue dotted line), and permanent electric field ($E_{MXY}$, red dotted line) for (a) MoSSe/Gr in Mode-II, (b) MoSTe/Gr in Mode-II, (c) MoSeTe/Gr in Mode-II, and (d) MoSeTe/Gr in Mode-I. The axis scale is the same in all panels.*

## Conclusions

In summary, the structural, electronic, and Schottky barrier properties of Janus heterostructures *MXY/Gr* (*M=Mo, W; X, Y=S, Se, Te*) have been systematically investigated by first principles calculations. The simulations indicate both *YMX/Gr* (*Mode-I*) and *XMY/Gr* (*Mode-II*) interfacing geometries are stable owing to a negative binding energy. Through varying the composition and interface geometry, the type of interface contact can be tuned from Ohmic to p-type or n-type Schottky with substantial changes of the Schottky barrier height in a $0.1 < \Phi_p < 0.7$ eV and $0.2 < \Phi_n < 0.6$ eV range. This unique property stems from the interplay between the permanent MXY electrostatic dipole and the induced one caused by charge redistribution at the MXY/Gr interface. Such an interplay is shown to be highly effective in altering the electrostatic potential difference across the vdW Janus heterostructure, directly affecting the $E_{SBH}$ and Schottky (Ohmic) contact types. We believe our computational findings contribute valuable new insights and guidelines towards design and development of function-tailored electrical contacts in nanoscale devices.

## Acknowledgements

This work was supported by the Science Challenge Project (TZ2018004), the National Natural Science Foundation of China (Nos. 51572016, U1530401, 11704116, 11804090, 51472209, 11774298, U1401241, and 21503012), the Natural Science Foundation of Hunan Province, China (Grant No. 2019JJ50175 and 2019JJ50148). This research was also supported by a Tianhe-2JK computing time award at the Beijing Computational Science Research Center (CSRC). G. T. acknowledges support by the Royal Society Newton Advanced Fellowship scheme (grant No. NAF\R1\180242).